\documentclass[aps,prb, twocolumn,showpacs,preprintnumbers,superscriptaddress]{revtex4}

\usepackage[dvipdfm]{graphicx}
\usepackage{dcolumn}
\usepackage{bm}

\begin{document}

\title{Restoration of  magnetization reversal under hydrostatic pressure in the lightly electron-doped manganite compound (Ca,Sr)Mn$_{0.95}$Sb$_{0.05}$O$_{3}$}

\author{Takahiro Fujiwara}
\author{Michiaki Matsukawa} 
\email{matsukawa@iwate-u.ac.jp }
\author{Takahiro Aoyagi} 
\author{Satoru Kobayashi} 
\affiliation{Department of Materials Science and Engineering, Iwate University, Morioka 020-8551, Japan}
\author{Shigeki Nimori}
\affiliation{National Institute for Materials Science, Tsukuba 305-0047,Japan}
\author{Ramanathan Suryanarayanan}
\affiliation{Laboratoire de Physico-Chimie de L'\'{e}tat Solide, CNRS,UMR8182, Universit\'{e} Paris-Sud, 91405 Orsay,France}
\date{\today}

\begin{abstract}
We report on the anomalous magnetization reversal under hydrostatic pressure in the lightly electron-doped manganite compound (Ca$_{1-x}$Sr$_{x}$)Mn$_{0.95}$Sb$_{0.05}$O$_{3}$ with its fixed carrier content.  
In a weakly magnetic field cooled measurement, diamagnetic magnetization is observed for $x\leq 15\%$, which  changes to positive values for $x>15\%$.  However, on an application of pressure on  the samples with $x=16\%$ and 17$\%$,  magnetization reverses sign and diamagnetism is restored. We present a magnetic phase diagram as a function of Sr concentration,  under both ambient and hydrostatic pressures.   
To understand better the thermodynamical properties of this  system, we have measured  the specific heat as a function of temperature under a field cooling of 100 Oe.
Our data show no anomalies associated with the temperature dependent magnetization reversal, indicating the absence of a long-rang magnetic ordering. 
The $ac$ magnetic susceptibility measurement points to the existence  of magnetic frustration  for the  Sr substituted samples exhibiting diamagnetic behavior.

\end{abstract}

\renewcommand{\figurename}{Fig.}
\maketitle

Extensive studies on the magnetization reversal or negative magnetization in magnetic materials demonstrate  the potential use in magnetic memory and related applications. 
CaMnO$_{3}$, the end member of the Ca$_{1-x}$La$_{x}$MnO$_{3}$  system, undergoes a G-type antiferromagnetic transition around $T_N\sim 120$ K, accompanied by a weak ferromagnetic component,\cite{MAC67} where each spin of the Mn ions is antiparallel to the nearest neighbors of Mn. In recent years, the electron-doped manganite system ($x<0.5$)\cite{CH96} has attracted much attentions because of the possibility of observing a negative magnetoresistance effect similar to that observed in its counterparts, in the so-called hole-doped manganites, for $x>0.5$. 
The negative magnetization phenomena in manganites were originally reported in compounds with two sublattices of Mn ions and rare-earth ions (Nd, Gd, Dy) \cite{BAR05,PEN02,NO96}
Some of these studies were discussed on the basis of a ferrimagnetic scenario leading to a negative magnetization at temperatures below a compensation temperature, where Mn and some rare-earths sublattices are antiferrmagnetically coupled.  Earlier, a negative magnetization in CaMnO$_{3}$ with  B-site substitution had been reported\cite{ANG06}. 
Recently, we demonstrated the effect of the hydrostatic pressure on magnetic and transport properties in the lightly electron-doped manganite CaMn$_{1-x}$Sb$_{x}$O$_{3}$ \cite{MU11}.  Anomalous magnetization reversals were clearly observed for $x$=0.05 and 0.08  in the field-cooled magnetization while the application of external pressure induced a suppression of the negative magnetization. 
A theoretical work on lightly electron-doped manganite CaMnO$_{3}$  predicted that spin canting in the G-type antiferromagnetic structure is realized by electron doping through the double exchange mechanism.\cite{ZE51,OH12} The weak FM component observed in the Ce-substituted CaMnO$_{3}$ is well explained by the spin-canting G-type AFM state with the double-exchange hopping of electrons. 
For the lightly electron-doped samples with CaMn$_{1-x}$Sb$_{x}$O$_{3}$ ,  we believe that a weak ferromagnetic trend in their $MT$ curves is compatible with  the present canted AFM scenario.  The larger Sb substitution at the Mn site introduces the local lattice distortion of Mn$^{3+}$O$_{6}$ associated with e$_{g}$-electron doping  and  then changes the orbital state of e$_{g}$-electron through the local John-Teller effect. As a result, it  gives rise to a considerable variation in  local spin configuration existing at the nearest neighbor of its Sb ion, leading to the formation of diamagnetic clusters within its matrix. The canted AFM matrix stablized  by  light electron doping contributes to a weak ferromagnetic component. On the other hand,  the diamagnetic clusters are simultaneously formed through the local lattice deformation due to the substitution of Sb with its larger ionic radius.  
In this paper, we report   the magnetization reversal under hydrostatic pressure in the lightly electron-doped manganite compound (Ca$_{1-x}$Sr$_{x}$)Mn$_{0.95}$Sb$_{0.05}$O$_{3}$ with its fixed carrier content.  
Furthermore, to understand better the thermodynamical properties of the present system, we carry out the specific heat measurement  for these compounds over a wide range of temperature. 

\begin{figure}[ht]\includegraphics[width=6cm]{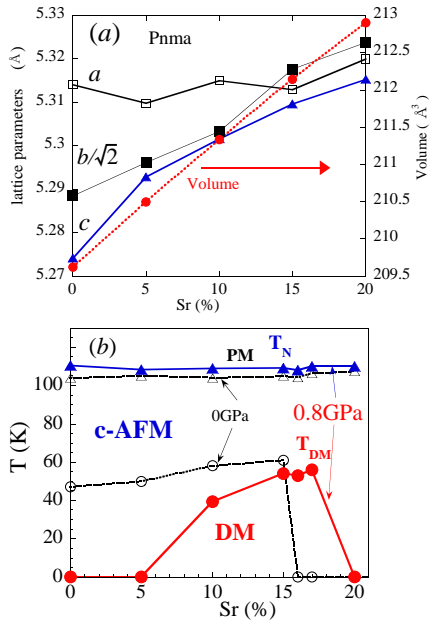}
\caption{(Color online) (a)The lattice parameters and unit-cell volume as a function of Sr  estimated from X-ray diffraction patterns of  (Ca$_{1-x}$Sr$_{x}$)Mn$_{0.95}$Sb$_{0.05}$O$_{3}$.
The unit-cell volume is nearly proportional to an increase of the  Sr content. (b) The magnetic phase diagram of the Sr-substituted (Ca$_{1-x}$Sr$_{x}$)Mn$_{0.95}$Sb$_{0.05}$O$_{3}$, where 
PM, c-AFM, and DM denote paramagnetic, canted antiferromagnetic, and diamagnetic phases, respectively.  $T_{DM}$ and  $T_{N}$  represent  the diamagnetic and canted antiferromagnetic transition temperatures. 
}\label{LA}
\end{figure}

Polycrystalline samples of (Ca$_{1-x}$Sr$_{x}$)Mn$_{0.95}$Sb$_{0.05}$O$_{3}$ ($x$=0.0, 0.05, 0.10, 0.15, 0.16, 0.17, and 0.2)  were prepared by using a solid-state reaction method. The stoichiometric mixtures of high purity CaCO$_{3}$, Mn$_{3}$O$_{4}$, Sb$_{2}$O$_{3}$ and SrCO$_{3}$  powders were calcined in air at 
$1000^{\circ}$C for 24 h. The products were then  ground and pressed into cylindrical pellets.  The pellets were finally sintered  at $1400\sim 1450^{\circ}$C for 12 h. 
X-ray diffraction data obtained by using  Rigaku Ultima IV revealed that all samples had  almost a single phase with an orthorhombic structure ($Pnma$)\cite{PO04}. 
The lattice parameters and unit-cell volume estimated from X-ray diffraction patterns using RIETAN-FP program are shown as a function of Sr content  in Fig. \ref{LA},
We note  the linear relationship between the volume and  the Sr substitution at the Ca site in present system, as  previously reported in  (Gd$_{0.08}$Ca$_{1-x}$Sr$_{x}$)MnO$_{3}$.\cite{ HI04}
X-ray photoemission spectroscopy on the  CaMn$_{0.8}$Sb$_{0.2}$O$_{3}$ sample performed at room temperature  strongly support that the valence of the Sb ion is 5+ .\cite{FU13}
The electron probe micro analyzer (EPMA) analysis on  the nominal samples  with $x$=0.0, 0.15, 0.16, and 0.17  is listed in Table \ref{table1}.   We note that  our samples prepared by the solid state reaction method are close to the nominal composition  and there is a quite small difference in the Mn content between each sample. 
The $dc$ magnetization measurement was carried out using the commercial superconducting quantum interference device (SQUID) magnetometers  at Iwate University and at the National Institute for Materials Science (NIMS). The $ac$ magnetic susceptibility measurement was measured  as a function of temperature for  frequencies ranging from 1 Hz to 1000 Hz under the  $ac$ magnetic field of 5 Oe at NIMS.  
The specific heat measurement in field cooled mode  was carried out   from 120 K down to 2 K using the physical property measuring system (Quantum Design PPMS). 

\begin{table}
\caption{\label{table1} EPMA  analysis  of  (Ca$_{1-x}$Sr$_{x}$)Mn$_{0.95}$Sb$_{0.05}$O$_{3}$}  ($x$=0.0, 0.15, 0.16, and 0.17) .
\begin{ruledtabular}
\begin{tabular}{ccccccccc}
Element&Ca&Sr& Mn&Sb &O&&&\\
nominal composition& &&&&&&&\\
\hline
 CaMn$_{0.95}$Sb$_{0.05}$O$_{3}$ &1.010&0.00\ &0.943&0.048&3.02&&\\
(Ca$_{0.85}$Sr$_{0.15}$)Mn$_{0.95}$Sb$_{0.05}$O$_{3}$&0.845 &0.164&0.927&0.048&3.03&&\\
(Ca$_{0.84}$Sr$_{0.16}$)Mn$_{0.95}$Sb$_{0.05}$O$_{3}$&0.836&0.177&0.925&0.048&3.02&& \\
(Ca$_{0.83}$Sr$_{0.17}$)Mn$_{0.95}$Sb$_{0.05}$O$_{3}$&0.830&0.187&0.918&0.047&3.04&&\\
\end{tabular}
\end{ruledtabular}
\end{table}

\begin{figure}[ht]\includegraphics[width=7cm]{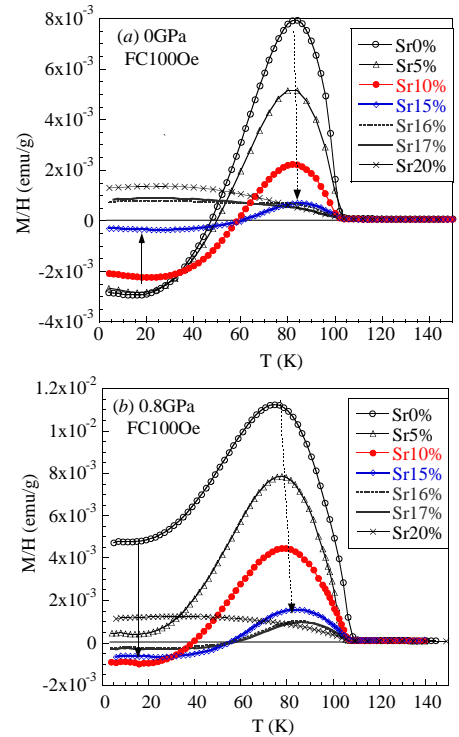}
\caption{(Color online) Temperature dependence of the field-cooled  magnetic susceptibility $M/H$  for  (Ca$_{1-x}$Sr$_{x}$)Mn$_{0.95}$Sb$_{0.05}$O$_{3}$ recorded under both  ambient and hydrostatic pressures, (a) and (b). 
}\label{MT}
\end{figure}%

\begin{figure}[ht]\includegraphics[width=7cm]{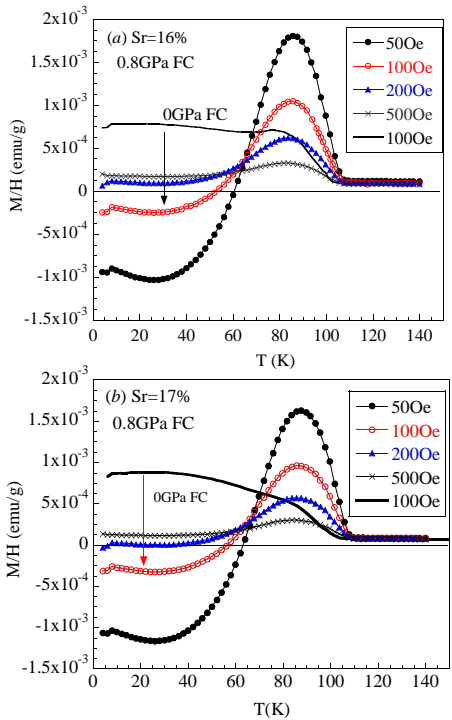}
\caption{(Color online) Temperature dependence of the field-cooled  magnetic susceptibility $M/H$  for  (Ca$_{1-x}$Sr$_{x}$)Mn$_{0.95}$Sb$_{0.05}$O$_{3}$ recorded under a hydrostatic pressure of 0.8 GPa.  (a) $x$=0.16, and (b)  $x$= 0.17. For comparison,  the solid curves represent  the  FC data without  the applied pressure. 
}\label{MT2}
\end{figure}%
\begin{figure}[ht]\includegraphics[width=8cm]{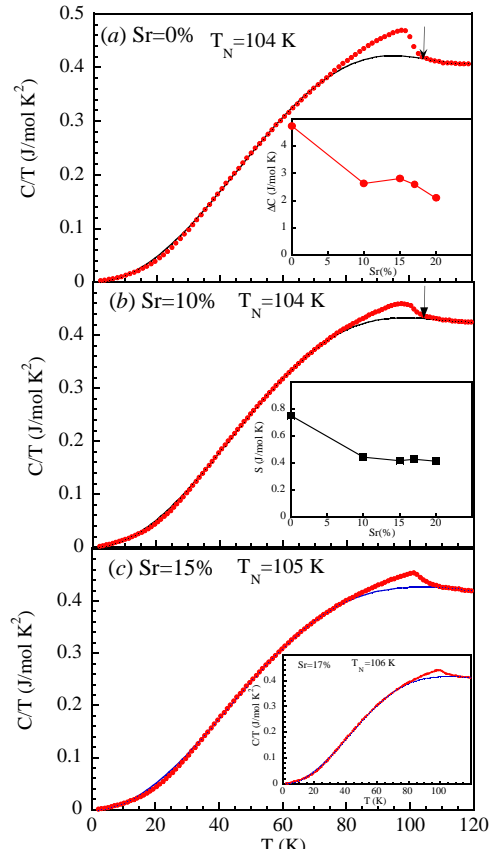}
\caption{(Color online)  $C/T$  vs $T$ for  (Ca$_{1-x}$Sr$_{x}$)Mn$_{0.95}$Sb$_{0.05}$O$_{3}$  measured in field cooling of 100 Oe .  (a) $x$=0, (b)$x$=0.1
and (c)$x$=0.15  (the inset of (c) $x$=0.17). The solid curves denote  the lattice contribution estimated from a polynomial approximation.  
The insets of (a) and (b) represent the magnetic heat jump $\Delta C$  and the magnetic entropy $S_{mag}$ associated with the magnetic transition,  respectively. 
}\label{CT}
\end{figure}%

\begin{figure}[ht]\includegraphics[width=7cm]{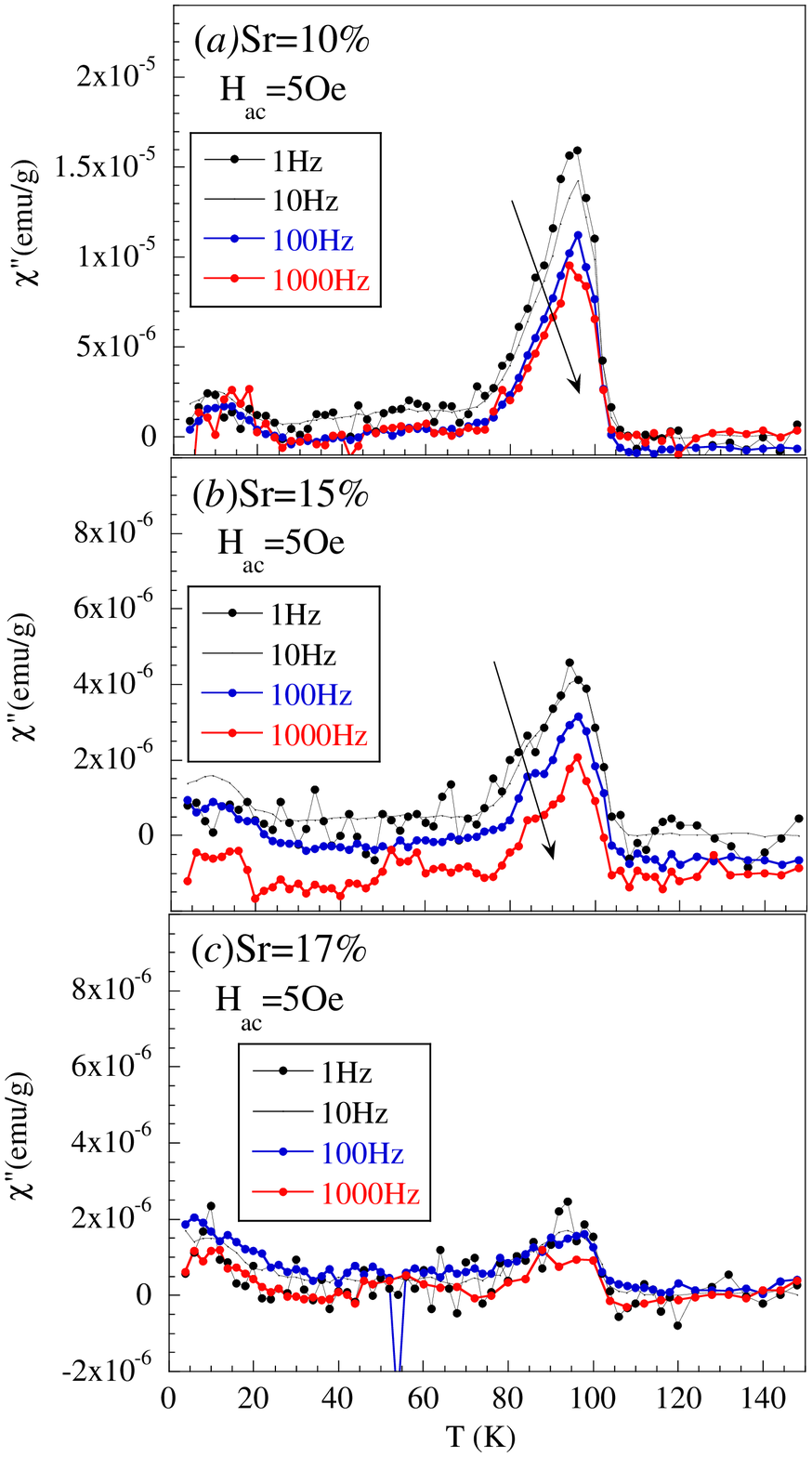}
\caption{(Color online) The imaginary parts of $ac$ magnetic susceptibility of (Ca$_{1-x}$Sr$_{x}$)Mn$_{0.95}$Sb$_{0.05}$O$_{3}$ as a function of temperature  collected at zero $dc$ magnetic field for frequencies  ranging from 1 Hz to 1000 Hz.  (a)  $x=0.10$, (b)  $x=0.15$, and (c) $x=0.17$.   The amplitude of the $ac$ magnetic field $H_{ac}$ was  5 Oe. The arrows point to the direction of increasing frequencies.  
}\label{AC}
\end{figure}%

Now, let us show in Fig.\ref{MT}  , the effect of  Sr substitution at Ca site on the temperature dependent magnetization reversal for  (Ca$_{1-x}$Sr$_{x}$)Mn$_{0.95}$Sb$_{0.05}$O$_{3}$.
For the parent sample, upon lowering temperature  the magnetization shows a large enhancement near 80 K  and  then the diamagnetic behavior is observed  at lower temperatures.   
The Sr substitution at Ca site affects  a strong suppression in the magnetic peak located around  80 K from  $8\times 10^{-3}$ emu/g at  $x$=0.0 through  $5\times 10^{-3}$ emu/g at  $x$=0.05 down to $2\times 10^{-3}$ emu/g at  $x$=0.10.  On the other hand,  a negative value  of  the magnetization at 20 K is not largely changed  in the case of Sr content up to $x$=0.10.  
 The effect of  Sr substitution on the  crystal structures of  perovskite manganites   (Ca$_{1-x}$Sr$_{x}$)MnO$_{3}$  and   (Gd$_{0.08}$Ca$_{1-x}$Sr$_{x}$)MnO$_{3}$ brings  a suppression of the Mn-O-Mn buckling.\cite{CH01,HI04}  It is pointed out in lightly electron doped manganites\cite{HI04} that the unit-cell volume increases linearly with an increase of  the Sr substitution at the Ca site, which  is caused by not the expansion of Mn-O bond length but a suppression of  the tilting of the MnO$_{6}$  octahedra.    Our lattice data  also exhibit  the linear relationship between the volume and Sr content  as shown in Fig.\ref{LA} (a). 
In the case of  low Sr,   the Sr ions are distributed within the canted AFM matrix contributing to weak ferromagnetic component.   If the Sr substitution at Ca site  gives rise to a suppression of the buckling of Mn-O-Mn bonds,  it then results in a suppression of the magnetic peak.  
A further increase of  Sr ions  above $10\%$ affects  diamagnetic clusters occupying  small regions around Sb site  and at last  the addition of Sr  beyond 15$\%$  suppresses the diamagnetic magnetization . 
We expect that   the tilting of  Mn$^{3+}$O$_{6}$  octahedron  near  Sb ion  is relaxed  by  the higher Sr substitution,  giving a decrease in the negative component. 
It is an interesting feature that   the disappearance of the negative magnetization accompanies a collapse  of the clear peak in the magnetization curve.  


Next,  we show in Fig.\ref{MT} (b)   the magnetization curve under the applied pressure of 0.8 GPa, to examine further the lattice effect on the magnetic properties  of (Ca$_{1-x}$Sr$_{x}$)Mn$_{0.95}$Sb$_{0.05}$O$_{3}$.
Application of the hydrostatic pressure on  the parent sample enhances  the magnetic susceptibility  over a wide range of temperatures,  the magnitude of  $M/H$ at 80 K increases from $8\times 10^{-3}$  emu/g at 0 GPa   to  $1.1\times 10^{-2}$  emu/g at 0.8 GPa, and  the diamagnetic behavior at lower temperatures is not visible.
Upon increasing Sr content up to  $15\%$,  the magnetic peak is considerably suppressed as observed in the case of ambient pressure. 
However,   the low temperature magnetization curve recorded at 0.8 GPa  is changed  from  positive values  at  Sr=$0\%$ and $5\%$ to negative values at  Sr=$10\%$.
There exist  qualitative differences in the Sr dependence of  the low-$T$  magnetization between both  ambient and  hydrostatic pressures. 
 These findings  are close to differences in  the response of the lattice between the chemical and physical pressure effects. 
In previous works on the effect of pressure on the atomic structure of manganites,  below 2 GPa  all three Mn-O bond lengths are compressed but the Mn-O bond angles show no obvious changes.\cite{CU03,CH10}
According to these studies,   the applied pressure on present samples causes a shrinkage of Mn-O bond lengths, resulting in  the enhancement of  the magnetic coupling  between Mn ions, which explains well 
both the steep increase of  $M/H$ and a stable rise of  the weak ferromagnetic transition temperature. 
For the samples with  $x=16\%$ and 17$\%$, the diamagnetism reappears  under 0.8 GPa in spite of  no indication of the negative magnetization under  the ambient pressure, as shown in Fig. \ref{MT2} .
The local JT distortion is more enhanced through the orbital lattice coupling  under the applied pressure in comparison to the pressure induced enhancement  of magnetic interaction, 
which is probably close to  the reappearance of canted spin clusters  contributing to a diamagnetism.  
The magnetic phase diagram of  (Ca$_{1-x}$Sr$_{x}$)Mn$_{0.95}$Sb$_{0.05}$O$_{3}$ established at 100 Oe under the ambient and hydrostatic pressures is summarized in the inset of Fig.\ref{LA} (b).

To understand further  the thermodynamical properties of this  system, we  performed  the specific heat measurements over a wide range of temperatures  under a field cooling of 100 Oe.
Figure \ref{CT} shows the specific heat  as a function of temperature from 2 K to 120 K for  (Ca$_{1-x}$Sr$_{x}$)Mn$_{0.95}$Sb$_{0.05}$O$_{3}$  ( $x$=0, 0.1, 0.15, and 0.17) .
Here, the lattice contribution $C_{l}$ is roughly estimated from a polynomial approximation as previously reported in manganites. \cite{GO98}
$\Delta C$  represents the maximum value of  the magnetic contribution  $C-C_{l}$  and  the magnetic entropy is estimated by integrating  $(C-C_{l})/T$  with respect to $T$. 
We observe clear anomalies associated with a magnetic transition  near $T_{N}$ but  no visible changes in  the specific heat  data at temperatures where  the diamagnetic behaviors appear. 
Our thermodynamical data strongly support  that  the temperature dependent magnetization reversal  accompanies no long-rang magnetic ordering. 
It is  expected from theses findings  that  short-range magnetic clusters around Sb ions have a close relationship with  the appearance of observed weak diamagnetism. 
For the Sr substituted samples,  both  the magnetic heat jump and magnetic entropy are not largely changed. 
In particular, we do not find out  distinct differences in the thermodynamical properties between   the diamagnetic Sr15$\%$ sample  and the weakly ferromagnetic  Sr17$\%$  one. 

Finally, we carried out the $ac$  magnetic susceptibility measurements for (Ca$_{1-x}$Sr$_{x}$)Mn$_{0.95}$Sb$_{0.05}$O$_{3}$ ($x$ 0.1, 0.15, and 0.17), in order to clarify the dynamical effect on the diamagnetic state. The temperature dependence of the imaginary parts, $\chi''$, are registered at zero $dc$ magnetic field with increasing frequency $f$ ranging from 1 Hz to 1000 Hz , as shown in Figs. \ref{AC}.
 For the parent Sr0$\%$ sample, a sharp peak in  $\chi''$ is observed around 100 K, which is good agreement with the $dc$ magnetic measurement. \cite{MU11}
Upon increasing $f$,  a second peak of $\chi'$ and $\chi''$ located at 90 K slightly shifts towards lower temperatures. For lower Sr doping up to 10$\%$, the amplitude of $\chi''$ shows its strong decay with increasing frequency. When the Sr content exceeds 15$\%$, the frequency dependence of $ac$ magnetic susceptibility almost disappears. The frequency dependence of the $ac$ magnetization indicates the signature of a spin-glass like character, but a substantial decrease of the peak in $\chi''$ with increasing $f$ is in contrast to the behavior of conventional spin-glass system \cite{MU81}, as previously pointed out in the phase-separated  Pr$_{0.7}$Ca$_{0.3}$MnO$_{3}$.\cite{DE01} 
The $ac$ magnetic susceptibility measurement  of the lower Sr substituted samples suggests  the existence of magnetic frustration between diamagnetic clusters and canted antiferromagnetic matrix. 
In summary, we have demonstrated the remarkable influence of pressure on the magnetic properties of the title compound. For  $x>15\%$,  the disappeared diamagnetic signal, reappears when the pressure is applied. However, the specific heat measurements do not reveal any magnetic phase transition. The  $ac$ susceptibility data indicate the existence of magnetic frustration.

\end{document}